\documentclass[oldversion]{aa}  
\usepackage{graphicx}
\usepackage{txfonts}

\def\pro2{\textsc{\bfseries pro2}}
\def\lte2{\textsc{\bfseries lte2}}
\def\atoms2{\textsc{\bfseries atoms2}}
\def\setf2{\textsc{\bfseries setf2}}
\def\line1prof{\mbox{\textsc{\bfseries line1}\raisebox{-0.5ex}{\bfseries--}\textsc{\bfseries prof}}}

\def\etal{{et\,al.}\ }

\def\keins{K1-16}
\def\hh{H1504$+$65}
\def\rxj{RX\,J2117.1$+$3412}
\def\ngc{NGC\,246}
\def\lovier{Longmore\,4}
\def\kpd{KPD\,0005+5106}

\newcommand{\Teff}{$T\mathrm{\hspace*{-0.4ex}_{eff}}$\,}
\newcommand{\logg}{$\log\,g$\hspace*{0.5ex}}

\newcommand{\gppr}{\stackrel{>}{\scriptstyle \sim}}
\newcommand{\gappr}{\raisebox{-0.4ex}{$\gppr $}}

\begin{document}
\title{Discovery of photospheric Ca\,X emission lines in the far-UV spectrum
  of the hottest known white dwarf (\kpd)\thanks{Based on observations
  made with the NASA-CNES-CSA \emph{Far Ultraviolet Spectroscopic
  Explorer}. \emph{FUSE} is operated for NASA by the Johns Hopkins
  University under NASA contract NAS5-32985.}  }

\author{K\@. Werner\inst{1}
   \and T\@. Rauch\inst{1}
   \and J.~W\@. Kruk\inst{2}}

\institute{Institut f\"ur Astronomie und Astrophysik, Kepler Center for Astro and Particle Physics, 
Eberhard-Karls-Universit\"at, Sand 1, 72076 T\"ubingen, Germany. \email{werner@astro.uni-tuebingen.de}
   \and Department of Physics and Astronomy, Johns Hopkins University, Baltimore, MD 21218, USA}

\date{Received 13 October 2008 / Accepted 7 November 2008}

\authorrunning{K. Werner \etal}
\titlerunning{\ion{Ca}{X} emission lines in \kpd}

\abstract{For the first time, we have identified photospheric emission
  lines in the far-UV spectrum of a white dwarf. They were discovered in
  the \emph{Far Ultraviolet Spectroscopic Explorer} spectrum of the hot
  (\Teff\,$\approx$\,200\,000\,K) DO white dwarf \kpd\ and they stem
  from extremely highly ionized calcium (\ion{Ca}{X}
  $\lambda\lambda$~1137, 1159\,\AA). Their photospheric origin is
  confirmed by non-LTE line-formation calculations. This is the highest
  ionisation stage of any element ever observed in a stellar
  photosphere. Calcium has never been detected before in any hot white
  dwarf or central star of planetary nebula. The calcium abundance
  determination for \kpd\ (1-10 times solar) is difficult, because the
  line strengths are rather sensitive to current uncertainties in the
  knowledge of effective temperature and surface gravity.   We discuss
  the possibility that the calcium abundance is much lower than expected
  from diffusion/levitation equilibrium theory.  The same  emission
  lines are exhibited by the [WCE]-type central star NGC\,2371. Another
  \ion{Ca}{X} line pair ($\lambda\lambda$~1461, 1504\,\AA) is probably
  present in a \emph{Hubble Space Telescope} spectrum of the PG1159-type
  central star NGC\,246.}

\keywords{Stars: abundances -- 
          Stars: atmospheres -- 
          Stars: evolution  -- 
          Stars: AGB and post-AGB --
          White dwarfs}

\maketitle
%
%________________________________________________________________

\begin{figure*}[tbp]
\begin{center}
\includegraphics[width=0.80\textwidth]{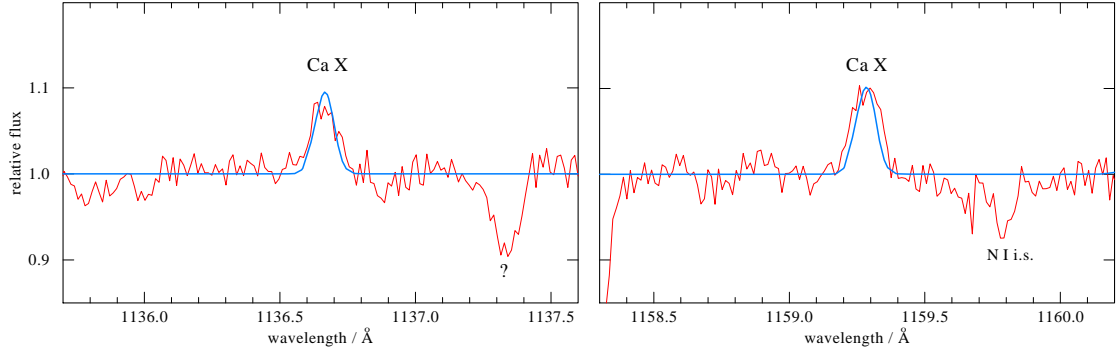}
  \caption[]{The two \ion{Ca}{X} emission lines discovered in \kpd\ (thin graphs). Overplotted is the spectrum from a model with
    \Teff=\,200\,000\,K, \logg=\,6.2, and solar Ca abundance. The model was convolved with a Gaussian
    with FWHM=0.05\,\AA\ in order to match the instrumental resolution.  \label{fig_kpd}
}
  \label{fig_all}
\end{center}
\end{figure*}

\section{Introduction}
\label{intro}

Observations of hot (pre-) white dwarfs with the \emph{Far Ultraviolet
Spectroscopic Explorer} (\emph{FUSE}) have revealed a large number of
chemical elements that were never detected before in these
objects. Their abundances can be used either to probe interior processes
in previous stellar evolution phases or to test predictions from
theories for element diffusion and radiative levitation.

We have recently identified \ion{Ne}{viii} lines in the hottest
(\Teff\,$\gappr$\,150\,000\,K) non-DA (pre-) white dwarfs,
i.e\@. objects of spectral type PG1159, DO, and [WCE] (Werner \etal
2007). The discovery of these lines in the hottest known DO white dwarf
\kpd\ was particularly surprising, because this proves that its
effective temperature must be much higher than previously thought
(200\,000\,K instead of 120\,000\,K).

\kpd\ was frequently observed by \emph{FUSE}  as a calibration target
over its entire lifetime. We have co-added all available spectra
and obtained datasets with
very high S/N ratio. A careful inspection of spectra taken with
different detectors revealed the presence of two hitherto unidentified
\emph{emission} lines. While there is still a large number of
unidentified absorption lines present in \emph{FUSE} spectra of hot
white dwarfs (WDs),  the discovery of emission features is unique and was
completely unexpected. In this \emph{Letter} we identify them as
\ion{Ca}{X} lines and present results of non-LTE modeling in order to
confirm their photospheric origin and to perform an abundance
determination.

We present observations and line identifications in
Sect.\,\ref{observations} and describe the modeling in
Sect.\,\ref{modeling}. The results from line-profile fits are presented
in Sect.\,\ref{results}. We conclude with Sect.\,\ref{conclusions}.

\section{Observations and \ion{Ca}{x} line identifications}
\label{observations}

The \emph{FUSE} instrument consists of four independent co-aligned
telescopes and spectrographs; two with Al+LiF optical coatings and two
with SiC coatings.  Taken together, the four channels span the
wavelength range 904--1187\,\AA\ with a typical resolving  power of
$R$\,$\approx$\,20\,000.  Further information on the \emph{FUSE} mission
and instrument can be found in Moos \etal (2000) and Sahnow \etal
(2000). \kpd\ was observed as a wavelength calibration object throughout
the \emph{FUSE} mission (observations M1070201--M1070234), in each of
the spectrograph apertures (LWRS, MDRS, HIRS), and once under program
P1040101 in LWRS. All observations were obtained in TTAG mode, except
for three early HIST mode observations; the latter were excluded as they
have different residual distortions in the wavelength scale. All
exposures were processed with the final version, v\,3.2.2, of CalFUSE
(Dixon \etal 2007). Spectra for each channel were shifted to place
absorption lines from low-ionization interstellar gas at a heliocentric
velocity of $-$15\,km\,s$^{-1}$ (Werner \etal 1996), and
combined. Variations in losses due to event bursts and channel
misalignments caused the net exposure times to vary: HIRS times varied
from 7.0\,ks in SiC2a to 17.9\,ks in LiF2b; MDRS times varied from
11.6\,ks in SiC1a to 18.6\,ks in LiF2b; LWRS times varied from 28.5\,ks
in LiF1b to 33.6\,ks in LiF2b. The net signal ranges from 700 counts per
0.013\,\AA\ pixel in the HIRS spectra to 1200\,c/pix in MDRS and as much
as 2700\,c/pix in LWRS. The \ion{Ca}{x} emission features are seen
clearly in each of the six available spectra (LiF1b and LiF2a in each of
HIRS, MDRS, LWRS). These six spectra were then resampled onto a common
wavelength scale and combined to produce a single spectrum with an
effective exposure time of 122.2\,ks.

The UV spectrum of NGC\,246, taken with the \emph{Space Telescope
Imaging Spectrograph} (\emph{STIS}) with grating E140H aboard the
\emph{Hubble Space Telescope} (\emph{HST}) was retrieved from the MAST
archive.

\begin{figure}[tbp]
\begin{center}
\includegraphics[width=0.85\columnwidth]{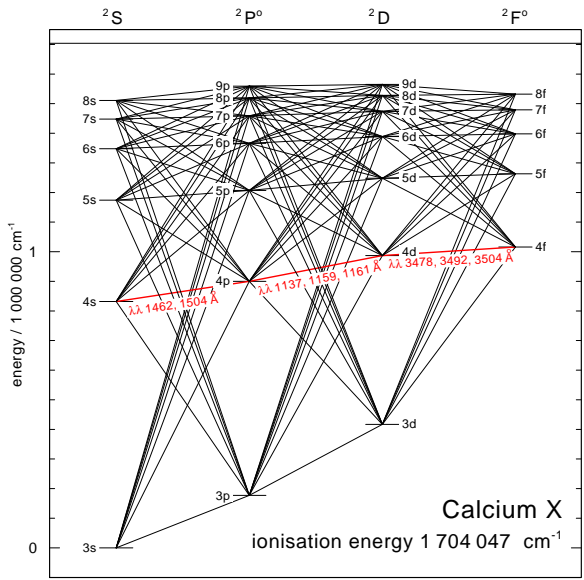}
  \caption[]{Grotrian diagram of our \ion{Ca}{x} model ion. 
Lines discussed in the text are caused by transitions between $n=4$
sublevels. The 4p--4d transition causes the observed UV emission lines.
\label{fig_grotrian}
}
\end{center}
\end{figure}

In the \emph{FUSE} spectrum of \kpd\ we detected two emission lines,
located at photospheric rest wavelengths $\lambda\lambda$~1136.5,
1159.2\,\AA\ (Fig.\,\ref{fig_kpd}; radial velocity $+$35\,km\,s$^{-1}$;
Werner \etal 1996). We identify these lines as due the
4p\,$^2$S\,--\,4d\,$^2$P$^{\rm o}$ transition in the \ion{Ca}{X} ion
(Fig.\,\ref{fig_grotrian}). Compared to the Ritz wavelengths in the
NIST\footnote{http://physics.nist.gov/ PhysRefData/ASD/index.html}
database, both observed lines are located at wavelengths shorter by
0.3\,\AA. Their NIST $gf$-values are $\log g_i f_{ik}=0.23$ and 0.46,
respectively. The third line component of this transition is located at
$\lambda$~1161.4\,\AA\ according to NIST, so that in reality we expect
it to be found at $\lambda$~1161.1\,\AA. Its $gf$-value, however, is
much smaller ($\log g_i f_{ik}=-0.47$) explaining the fact that we
cannot detect it in the observation. (This is confirmed by our
line-formation calculations.)

\begin{figure}[tbp]
\begin{center}
\includegraphics[width=0.9\columnwidth]{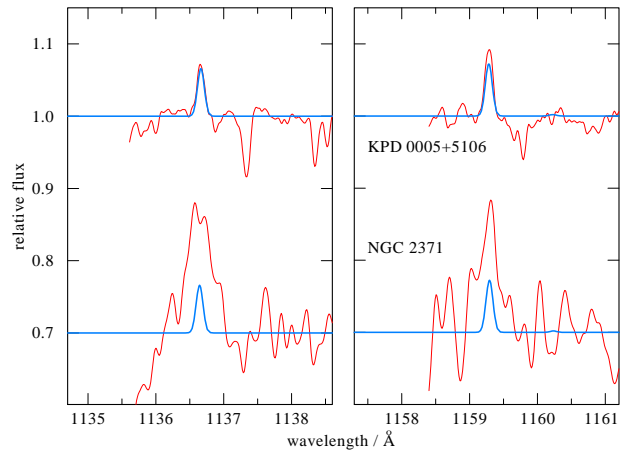}
  \caption[]{The two \ion{Ca}{x} emission lines discovered in \kpd\ (top spectra)
    are also seen in the [WCE] central star NGC\,2371. The model
    profiles are the same as in Fig.\,\ref{fig_kpd}. No attempt is made
    to fit the possibly wind-contaminated [WCE] profiles. For clarity, the
\emph{FUSE} spectra were smoothed with Gaussians (FWHM 0.05 and
0.1\,\AA, respectively). 
The model spectra (thick lines) were convolved with 0.1\,\AA\ Gaussians.
\label{fig_ngc2371}
}
\end{center}
\end{figure}

We searched for the two \ion{Ca}{X} lines in other hot DO white dwarfs
and PG1159 stars, but to no avail. As we will demonstrate below
(Sect.\,\ref{results}), this is a consequence of the extremely high
\Teff\ of \kpd\ combined with a relatively low surface gravity. However,
these lines are seen in the very hot, early-type Wolf-Rayet
central star NGC\,2371 (Fig.\,\ref{fig_ngc2371}). For this object we did
not attempt to fit these lines with our (static) model atmospheres,
because the profiles might be affected by the stellar wind.

\begin{figure}[tbp]
\begin{center}
\includegraphics[width=0.85\columnwidth]{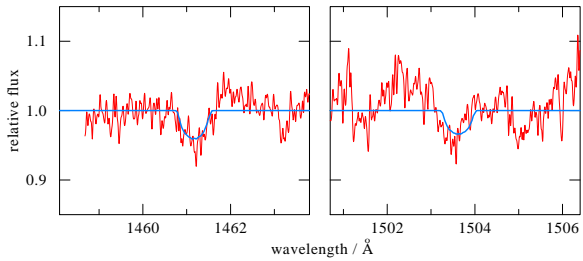}
  \caption[]{\emph{HST/STIS} spectrum of the PG1159-type
    central star NGC\,246 and computed profiles for the 4s--4p doublet
    of \ion{Ca}{X}. Their shape reflects the stellar rotation of
    $v$\,sin\,$i$\,=\,70\,km\,s$^{-1}$. The observation was smoothed with a Gaussian
    with FWHM=0.03\,\AA. 
\label{fig_ngc246}
}
\end{center}
\end{figure}

In the course of our model calculations we found that further
\ion{Ca}{X} lines might be detectable in other wavelength regions. The
4s--4p transition gives rise to a line doublet at
$\lambda\lambda$~1461.8, 1503.8\,\AA. Our models predict only marginal
emission features for \kpd, which cannot be detected in archival spectra
taken with the \emph{Faint Object Spectrograph} aboard \emph{HST} and
high-resolution spectra from the \emph{International Ultraviolet
Explorer}. However, for NGC\,246 our models predict absorption lines
that are possibly present in a \emph{HST/STIS} spectrum
(Fig.\,\ref{fig_ngc246}). The positions of the tentatively identified
absorption features in NGC\,246 differ from the NIST wavelengths by
$-0.7$ and $-0.2$\,\AA, respectively.

From NIST level energies one expects the two strongest lines of yet
another transition of \ion{Ca}{X} (4d--4f) to be located in the optical
UV at $\lambda\lambda$~3478, 3492\,\AA, respectively. A high-resolution
spectrum of \ngc\ taken with ESO's \emph{Very Large Telescope} and the
UVES spectrograph as part of the SPY survey (Napiwotzki \etal 2003)
reveals no line features there. Our model for NGC\,246 predicts
absorption lines with a depth of only 5\% relative to the continuum. The
relatively fast rotation ($v$\,sin\,$i$\,=\,70\,km\,s$^{-1}$) smears the
line features considerably, and they remain hidden in the noise.
Similar weak absorption profiles for these lines are predicted for \kpd,
but no appropriate observations are available.

\section{Model atmospheres and calcium line-formation}
\label{modeling}

We have designed a calcium model atom for NLTE line-formation
calculations. These were performed using and keeping fixed the physical
structure (temperature, densities) of line-blanketed NLTE model
atmospheres which are described in detail in Werner \etal (2004). In
short, they are plane-parallel and in hydrostatic and radiative
equilibrium. The models are composed of He, C, O, and Ne. For \kpd, we
assumed helium-dominated atmospheres with admixtures of C=0.003,
O=0.0006, Ne=0.01 (mass fractions). The high neon abundance was derived
from \ion{Ne}{viii} lines (Werner \etal 2007). The C and O abundances
are uncertain, because they were derived in earlier work that assumed
that \kpd\ is relatively cool (\Teff=\,120\,000\,K; Werner \etal
1996). We verified that varying the C and O abundances within reasonable
limits does not change the \ion{Ca}{X} lines significantly. A series of
models with various \Teff\ and \logg values was computed to study the
dependency of the \ion{Ca}{X} lines on these parameters (see
Sect.\,\ref{results}). For NGC\,246 we adopted \Teff=\,150\,000\,K,
\logg=\,5.7, and the composition He/C/O/Ne=0.62/0.30/0.06/0.02 (Werner
\etal 2007).

The Ca model atom considers the ionization stages {\scriptsize
VIII$-$XII}, represented by 1, 15, 25, 4, 1 NLTE levels, respectively,
plus a number of LTE levels. In the ions \ion{Ca}{ix$-$xi} we include
23, 126, and 2 line transitions, respectively.  Atomic data were taken
from the NIST, Opacity (Seaton \etal 1994), and IRON (Hummer \etal 1993)
Projects databases (TIPTOPbase\footnote{
http://cdsweb.u-strasbg.fr/topbase/}). Fine-structure splitting is
accounted for in the final formal solution for the synthetic
line-profile computation, distributing the level populations among
sublevels assuming LTE. For all lines we assumed quadratic Stark
broadening for the profile calculation.

\begin{figure}[tbp]
\begin{center}
\includegraphics[width=0.85\columnwidth]{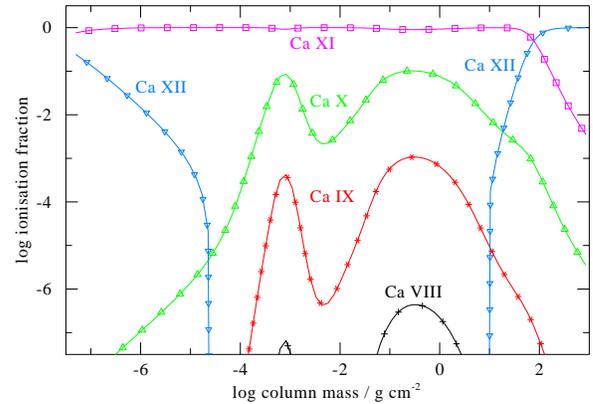}
  \caption[]{
Ionization fraction of calcium as a function of atmospheric
    depth in the model with \Teff=\,200\,000~K, \logg=\,6.2, and solar Ca
    abundance. 
\label{fig_ion}
}
\end{center}
\end{figure}

Particularly for the \ion{Ca}{X}~$\lambda\lambda$~1137, 1159\,\AA\ lines
the values of the oscillator strengths differ between the OP and NIST
databases. We use the OP values for our NLTE level population
iterations, because they are complete, in contrast to the NIST
database. For the final line-profile calculation we prefer the NIST
values, which are higher than the OP values by $\approx$\,25\%. The
differences do not affect our conclusions.

Photoionization cross-sections are taken from the Opacity Project
database when available or, otherwise, computed in a hydrogen-like
approximation. Electron collisional rates were calculated with the usual
approximation formulae. The Ca model atoms that were used for this
analysis have been developed in the framework of the \emph{German
Astrophysical Virtual Observatory}
(\emph{GAVO}\footnote{http://www.g-vo.org}) project and are provided
within the T\"ubingen Model-Atom Database
\emph{TMAD}\footnote{http://astro.uni-tuebingen.de/\raisebox{.2em}{\tiny
$\sim$}rauch/TMAD/TMAD.html}.

\section{Results}
\label{results}

Figure~\ref{fig_kpd} shows a fit to the \ion{Ca}{X} emission lines in
\kpd\ with a model \Teff=\,200\,000\,K, \logg=\,6.2, and a solar Ca
abundance ($\log\,{\rm Ca} = -4.22$, mass fraction; Asplund \etal
2005). In Fig.\,\ref{fig_ion} we show the ionization structure of Ca
throughout this model atmosphere. Within the entire line-forming region
\ion{Ca}{xi} is dominant, followed by \ion{Ca}{x}. 
In order to achieve
the observed emission strength in a model with this temperature and Ca
abundance, the surface gravity must be that low (\logg=\,6.2). This is
0.3~dex lower than what is preferred from the \ion{He}{ii} line spectrum
(Werner \etal 2007). We will show, however, that the fit to the
\ion{Ca}{X} lines can be achieved with more than one parameter set.

\begin{figure}[tbp]
\begin{center}
\includegraphics[width=1.0\columnwidth]{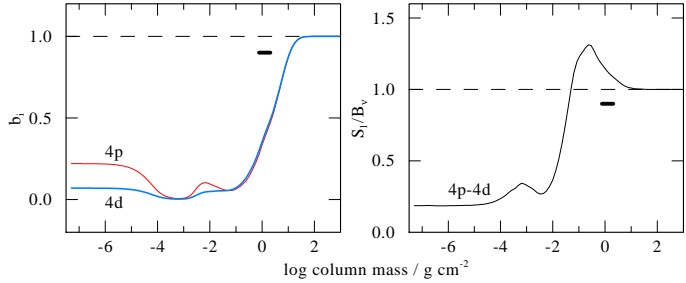}
  \caption[]{
\emph{Left:} Departure coefficients $b_{\rm i}$ for the 4p and 4d levels that cause
the \ion{Ca}{X} emission lines. \emph{Right:}  Ratio of 4p--4d
line source function $S_{l}$ to Planck function $B_\nu$ at line
centre. The thick horizontal lines near $\log m=0$ denote the
line formation region. Dashed lines correspond to the LTE case ($b_i=1$ 
and $S_{l}/B_\nu=1$). Model parameters as in Fig.\,\ref{fig_ion}.
\label{fig_depart}
}
\end{center}
\end{figure}

The occurrence of this line emission can be understood when the non-LTE
departure coefficients $b_i=n_i^{\rm NLTE}/n_i^{\rm LTE}$ for the
populations $n_i$ of the involved atomic levels and the line source
function $S_l$ are inspected (Fig.\,\ref{fig_depart}). The line source
function is determined by the ratio of the departure coefficients of the
lower and upper levels ($i,j$) and can be written as
$S_l/B_\nu=[\exp(h\nu_{ij}/kT)-1]/[(b_i/b_j)\exp(h\nu_{ij}/kT)-1])$,
where $B_\nu$ is the Planck function. An overpopulation of the upper
level relative to the lower (i.e. $S_l/B_\nu>1$) may lead to line
emission. In fact, this condition is fulfilled in the line-forming
region (Fig.\,\ref{fig_depart}), although both levels are underpopulated
(i.e. $b_i<1)$.

\begin{figure}[tbp]
\begin{center}
\includegraphics[width=\columnwidth]{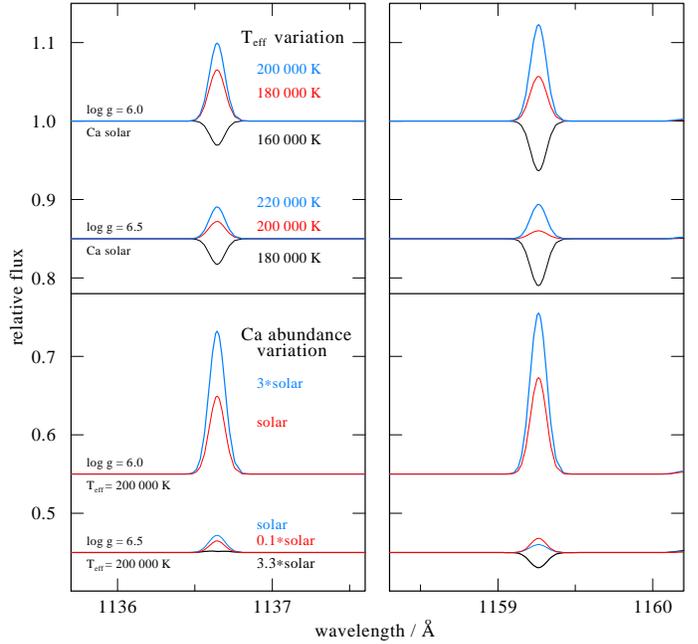}
  \caption[]{
Profile shapes of the \ion{Ca}{X} $\lambda\lambda$~1137, 1159\,\AA\ lines
  as a function of \Teff, \logg,
  and Ca abundance, as given by the labels.
\label{fig_models}
}
\end{center}
\end{figure}

We have computed a small grid of models with different \Teff\ and
\logg (representing the uncertainties with which these parameters are
known) and Ca abundances. The results are presented in
Fig.\,\ref{fig_models}.  Generally, high \Teff\ and low \logg is
necessary in order to bring this line into emission (left and right
upper panels). If \Teff\ decreases (and/or \logg increases), the lines
first turn from emission into weak absorption features and then
disappear at about 140\,000\,K. This explains why other DOs do not
exhibit these lines: they are significantly cooler and have higher
gravities compared to \kpd.

The sensitivity of these emission lines to the calcium abundance is
complicated and depends on \Teff\ and \logg of the
atmosphere (Fig.\,\ref{fig_models}, left and right lower
panels). Increasing the Ca abundance over the solar value can strongly
increase the emission height (model \Teff=\,200\,000\,K\, \logg=\,6) or
decrease it (model \Teff=\,200\,000\,K\, \logg=\,6.5). For \kpd\ we
achieved a good fit at \Teff=\,200\,000\,K, \logg=\,6.2, and solar Ca
abundance. As mentioned, however, our previous analysis favors a gravity
higher by 0.3 dex. Increasing the gravity to \logg=\,6.5 makes the
emission weaker, but this can be compensated by simultaneously
increasing the Ca abundance to 3 times the solar value and \Teff\ to
220\,000\,K. A much higher abundance, as predicted by diffusion theory, 
can be excluded. We have calculated models with 70 times solar
Ca abundance.  The emission line peak heights hardly change in the
\Teff=\,200\,000\,K, \logg=\,6.2 model when Ca is increased from solar
to 70 times solar; however, detailed inspection of the relative strength
of both lines shows that it is not in agreement with the observation. In
the observation as well as in the $\approx$\,solar Ca abundance models
the 1159\,\AA\ emission is stronger than the 1137\,\AA\ emission, as can
be expected from the $gf$-value ratio. In the 70 times solar models the
emission strength ratio is reversed, in contrast to the observation.

Concerning PG1159 stars, the behaviour of these Ca lines is rather
similar and therefore not shown in detail here. There are seven objects
that are hot enough and for which \emph{FUSE} spectroscopy is
available. These are the low-gravity central stars of planetary nebulae
\keins, \lovier, \rxj, \ngc\ (\Teff=\,140\,000--170\,000\,K,
\logg$\approx$\,5.5--6), the higher-gravity objects PG1520$+$525 and
PG1144$+$005 (\Teff=\,150\,000\,K, \logg=\,6.5--7), as well as the
peculiar \hh\ (\Teff=\,200\,000\,K, \logg=\,8). Model calculations were
performed for all of these objects with solar Ca abundance (diffusion is
not at work in these objects' atmospheres; see Unglaub \& Bues
2000). They exhibit the \ion{Ca}{x} $\lambda\lambda$~1137, 1159\,\AA\
lines as weak absorption features with a maximum depth of 5\% of the
continuum flux. Such weak lines cannot be detected in the available
\emph{FUSE} spectra.

\section{Summary and conclusions}
\label{conclusions}      

We have discovered \ion{Ca}{X} emission lines in far-UV spectra of the
DO white dwarf \kpd. This is the first detection of photospheric
emission lines in this spectral range of any hot \mbox{(pre-)} white
dwarf.  Provencal \etal (2005) discovered low-ionisation emission lines
in \emph{HST/STIS} UV spectra of two relatively cool
(\Teff$\approx$\,12\,500\,K) He-rich white dwarfs (spectral type DQ). It
was shown, however, that they are chromospheric in origin. The
\ion{Ca}{X} lines are the highest ionisation stage of any element
identified in any stellar photosphere. Our discovery also represents the
first identification of calcium in hot (pre-) WDs.

The Ca abundance in \kpd\ is in the range \mbox{$\approx$1--10} times
solar. A more precise determination from the emission lines is not
possible. A comparison of this result with predictions from radiative
levitation/gravitational diffusion equilibrium theory is difficult
because \Teff\ and \logg\ of \kpd\ are outside of the range considered
by Chayer \etal (1995; their Fig.\,20). For the closest parameters
(\Teff\,=\,130\,000\,K, \logg=7) a huge overabundance is predicted (2500
times solar). Our estimate for \logg\ is smaller (6.2--6.5) which would
result in an even higher overabundance. On the other hand it is
impossible to make a solid estimate for the effect of the higher \Teff\
(200\,000--220\,000\,K) on the behaviour of the Ca equilibrium
abundance, because the dominant ionisation stage in \kpd\ is
\ion{Ca}{xi}, while it is \ion{Ca}{viii} in the hottest Chayer \etal
model (\Teff\,=\,130\,000\,K, \logg=7.5). Looking at the behaviour of
other elements (S, Ar), namely how their equilibrium abundance changes
when their (respective isoelectronic) ionisation stages increase (with
increasing \Teff), it is suggestive that the Ca abundance at
\Teff\,=200\,000\,K is lower than at \,130\,000\,K, but not by orders of
magnitude. Although  detailed calculations are required for a definitive
statement, we conclude that the atmosphere of \kpd\ is probably not in
levitation/diffusion equilibrium.  This is confirmed by the
diffusion/mass-loss calculations of Unglaub \& Bues (2000) which suggest
that \kpd\ has yet to cross the wind-limit on its evolutionary track,
meaning that mass-loss is large enough to prevent both gravitational
settling and the accumulation of radiatively supported heavy
elements. In this case, \kpd\ is not a descendant of the PG1159
stars. An evolutionary link to the He-dominated central stars of
spectral type O(He) and to the RCrB stars has been suggested (Werner
\etal 2008).

If unaffected by diffusion processes, then the photospheric composition
of \kpd\ is the consequence of previous evolutionary phases. In
contrast, the presence of Ca in the atmospheres of cooler white dwarfs
(spectral types DAZ and DBZ, with low-ionisation optical Ca absorption
lines) requires on-going accretion of circumstellar matter, because
gravitational settling rapidly removes heavy elements from the
photosphere (e.g.\ Koester \& Wilken 2006).

The non-detection of the \ion{Ca}{x} $\lambda\lambda$~1137, 1159\,\AA\
lines in the hottest PG1159 stars is explained by undetectably weak
absorption line features in the models. Another \ion{Ca}{x} line pair
($\lambda\lambda$~1462, 1504\,\AA) is possibly present in absorption in
\ngc\ and suggests a roughly solar Ca abundance.  The only other object
in which we discovered the \ion{Ca}{x} $\lambda\lambda$~1137, 1159\,\AA\
emission lines is the [WCE]-type central star NGC\,2371. This
corroborates the extraordinarily high effective temperature of this
object.

\begin{acknowledgements} T.R. is supported
by the \emph{German Astrophysical Virtual Observatory} (\emph{GAVO}) project of the
German Federal Ministry of Education and Research under grant
05\,AC6VTB. J.W.K. is supported by the \emph{FUSE} project, funded by NASA
contract NAS5-32985. Some of the data presented in this paper were obtained from the
  Multimission Archive at the Space Telescope Science Institute
  (MAST).
\end{acknowledgements}

\end{document}